\documentclass{article}
\usepackage{spconf,amsmath,graphicx,bbm,mathtools,multirow,hhline}

\DeclareMathOperator*{\argmax}{arg\,max}

\title{Internal Language Model Estimation for Domain-Adaptive End-to-End Speech Recognition}
\name{\begin{tabular}{c}Zhong Meng, Sarangarajan Parthasarathy, Eric Sun, Yashesh Gaur, Naoyuki Kanda, \\ Liang Lu, Xie Chen, Rui Zhao, Jinyu Li, Yifan Gong\end{tabular}}
\address{Microsoft Corporation, Redmond, WA, USA}
\begin{document}
\ninept
\maketitle
\begin{abstract}
The external language models (LM) integration remains a challenging task for end-to-end (E2E) automatic speech recognition (ASR) which has no clear division between acoustic and language models.
In this work, we propose an internal LM estimation (ILME) method to facilitate a more effective integration of the external LM with all pre-existing E2E models with no additional model training, including the most popular recurrent neural network transducer (RNN-T) and attention-based encoder-decoder (AED) models. Trained with audio-transcript pairs, an E2E model implicitly learns an internal LM that characterizes the training data in the source domain. With ILME, the internal LM scores of an E2E model are estimated and subtracted from the log-linear interpolation between the scores of the E2E model and the external LM. The internal LM scores are approximated as the output of an E2E model when eliminating its acoustic components. ILME can alleviate the domain mismatch between training and testing, or improve the multi-domain E2E ASR.
Experimented with 30K-hour trained RNN-T and AED models, ILME achieves up to 15.5\% and 6.8\% relative word error rate reductions from Shallow Fusion on out-of-domain LibriSpeech and in-domain Microsoft production test sets, respectively. 



\end{abstract}
\section{Introduction}
\label{sec:intro}

End-to-end (E2E) automatic speech recognition (ASR), with the goal of directly mapping input speech features to output token sequence, has achieved state-of-the-art performance on a variety of tasks \cite{chiu2018state, sainath2020streaming, li2020developing}. 
The most popular E2E models include connectionist temporal classification \cite{graves2006connectionist, graves2014towards, hannun2014deep, soltau2016neural, Li18CTCnoOOV}, recurrent neural network transducer (RNN-T) \cite{graves2012sequence, jain2019rnn, sainath2020streaming,  li2020developing}, and attention-based encoder-decoder (AED) models \cite{chorowski2015attention, bahdanau2016end, chan2016listen, chiu2018state, karita2019comparative, Li2020Comparison}.
However, incorporating acoustic model (AM), language model (LM), and pronunciation models in a single deep neural networks (DNN) results in E2E models being more susceptible to domain shift from training to testing than conventional DNN-hidden Markov model (HMM) hybrid models \cite{DNN4ASR-hinton2012}. Many methods have been proposed to adapt ASR models towards the target domain of the test data, such as approaches based on regularization \cite{kld_yu, meng2019asa,l2_liao, meng2020lvector, lhuc, meng2019asa}, teacher-student learning \cite{li2014learning, meng2018adversarial, manohar2018teacher, meng2019conditional}, transformation \cite{lhn, tan2015cluster, sc_abdel, fhl}, and adversarial learning \cite{grl_shinohara, meng2018speaker, grl_serdyuk, dsn_meng}. However, all these methods require audio or audio-transcript pairs as the adaptation data when applied to E2E models \cite{ochiai2018speaker, meng2019speaker, meng2019domain}.

LM integration is a promising approach to adapt (customize) E2E models towards testing scenarios with \emph{text-only} data for both cross-domain and intra-domain applications. For cross-domain application, an external LM trained with target-domain text-only data is integrated with a source-domain E2E model to improve the ASR on the target-domain test data. For intra-domain application, an external LM trained with a large amount multi-domain text is fused with a multi-conditional E2E model to improve the ASR on multiple in-domain test data.
With orders of magnitude more text-only data available than E2E training transcripts, the external LM has great potential to correct the hypotheses of E2E models given unseen styles of speech or long tails of phrases and words in the test utterances. However, LM integration remains to be a challenging task because no explicit separation of AM and LM exists in E2E models.

Among many studies in LM integration, Shallow Fusion \cite{hannun2014deep, gulcehre2015on, chorowski2016towards, kannan2018analysis} stands as a simple yet effective approach involving a log-linear interpolation between the scores of the E2E model and a separately-trained LM during decoding. 
Towards a more structural integration, Deep Fusion \cite{gulcehre2015on}, Cold Fusion \cite{sriram2017cold, toshniwal2018comparison}, Simple Fusion \cite{stahlberg2018simple} and Component Fusion \cite{shan2019component} were proposed to jointly train an E2E model with an external LM to learn a sophisticated combination between their hidden units via gating mechanisms. However, with additional training steps and computational cost during decoding, these methods have not replaced Shallow Fusion as the dominant approach in LM integration. 

Recently, the Density Ratio method \cite{mcdermott2019density, kanda2016maximum, hori2017multi, kanda2017maximum} was proposed as an extension of Shallow Fusion. During inference, the score of a source-domain LM trained with E2E training transcript is subtracted from the log-linear combination of E2E model and LM scores. Density Ratio has shown to consistently outperform Shallow Fusion in a cross-domain evaluation. However, Density Ratio is based on the assumption that the E2E posterior is factorizable into an AM and an LM with individual parameters like a hybrid system, while the accurate factorization is to condition both the ``AM'' and ``LM'' components on the E2E model parameters. We call the token sequence probability predicted by an E2E model the \emph{internal LM}.

In \cite{variani2020hybrid}, a hybrid autoregressive transducer (HAT) was proposed as a new type of E2E model that preserves the modularity of the hybrid model. 
HAT provides an efficient way to estimate its internal LM by removing the effect of the encoder from the network. During HAT decoding, the internal LM score is subtracted from the path posterior after combining with external LM score. 
However, this LM integration requires us to train an HAT model with a special network architecture that separately models the label and duration distributions, preventing us from leveraging the existing E2E models.

Inspired by HAT, we propose an \emph{internal LM estimation (ILME)} method to effectively integrate external LMs with all pre-existing E2E models without any additional model training, including the most popular time-synchoronrous RNN-T and time-asynchonrous AED. With ILME, we estimate and subtract the internal LM scores of an E2E model from the log-linear combination of the E2E and LM scores during inference. By removing the separable AM-LM assumption, ILME enables a more accurate and a lighter-weight external LM integration than Density Ratio with no additional source-domain LM running at the inference. We evaluate 30K-hour RNN-T and AED models on cross-domain LibriSpeech test sets and intra-domain Microsoft production test sets. ILME achieves up to 15.5\% and 6.8\% relative word error rate (WER) reductions from Shallow Fusion for cross- and intra-domain evaluations, respectively. 
Further, despite having 26.8\%-29.8\% fewer parameters, ILME consistently outperforms Density Ratio in terms of lower WER.

\section{Related E2E Models}
An E2E model predicts the posterior distribution $P(\mathbf{Y} |
\mathbf{X};\theta_\text{E2E})$ over sequences of output tokens $\mathbf{Y}=\{y_1, \ldots,
y_U\}$ given a sequence of input speech features $\mathbf{X}=\{\mathbf{x}_1,
\ldots, \mathbf{x}_T\}$, where $y_u \in \mathcal{V}, u = 1, \ldots, U$, and $\mathbf{x}_t
\in \mathbbm{R}^{d_x}, t = 1, \ldots, T$. $\mathcal{V}$ is the set of all possible output tokens, e.g., word pieces, etc. For RNN-T and AED models, a special token $y_0=\texttt{<sos>}$ is inserted at the beginning of $\mathbf{Y}$ to indicate the start of the sentence.

\subsection{Recurrent Neural Network Transducer (RNN-T)}
The RNN-T model \cite{graves2012sequence} consists of an encoder, a prediction network and a joint network. Analogous to an acoustic model, the encoder maps the input speech features to a sequence of hidden representations $\mathbf{H}^{\text{enc}} = \{\mathbf{h}^\text{enc}_1, \ldots, \mathbf{h}^\text{enc}_T\}$, i.e., $\mathbf{H}^{\text{enc}} = \text{Encoder}(\mathbf{X})$. 
Imitating an RNN-LM, the prediction network takes the embedding vector $\mathbf{e}_{u-1}$ of the previous \emph{non-blank} token $y_{u-1}$ to generate the hidden representation $\mathbf{h}^\text{pred}_u$ by using an RNN, i.e., $\mathbf{h}^\text{pred}_u = \text{PredictionRNN}(\mathbf{h}^\text{pred}_{u - 1}, \mathbf{e}_{u - 1})$.

RNN-T predicts a conditional distribution over blank-augmented token sequences $\mathbf{\tilde{Y}} = \{\tilde{y}_1, \ldots, \tilde{y}_{T+U}\}$, where $\tilde{y}_i \in \mathcal{V}\: \cup\:\texttt{<b>}, i = 1, \ldots, T + U$, and $\texttt{<b>}$ is a blank. $\mathbf{\tilde{Y}}$ is aligned with the token and feature sequences $\mathbf{Y}$ and $\mathbf{X}$ as $\left(\tilde{y}_i, y_{u_i}, x_{t_i}\right)^{U + T}_{i=1}$,
where the index $i$ in $\mathbf{\tilde{Y}}$ is mapped to the index $u_i$ in $\mathbf{Y}$, and the index $t_i$ in $\mathbf{X}$.

The joint network combines the outputs of the encoder and prediction network via a feed-forward network to produce the logits $\mathbf{z}_{t_i, u_i}$, and applies a softmax to predict the conditional distribution over the next possible tokens in $\mathcal{V} \cup \texttt{<b>}$, i.e., 
\begin{align}
    &\mathbf{h}^\text{joint}_{t_i, u_i} = \phi(W_e\mathbf{h}^{\text{enc}}_{t_i} + W_p\mathbf{h}^\text{pred}_{u_i} + \mathbf{b}_e + \mathbf{b}_p), \label{eqn:joint} \\
    &\mathbf{z}_{t_i, u_i} = W_j \mathbf{h}^\text{joint}_{t_i, u_i} + \mathbf{b}_j, \label{eqn:rnnt_logit} \\
    & \hspace{-4pt} \left[P(\tilde{y}_i=v|\mathbf{X}_{1:t_i}, \mathbf{Y}_{0:u_{i - 1}}; \theta^\text{S}_\text{RNNT})\right]_{v\in \mathcal{V} \cup \texttt{<b>}} \hspace{-7pt}= \text{softmax}(\mathbf{z}_{t_i, u_i}), \label{eqn:rnnt_softmax}
\end{align}
where $\phi$ is a non-linear function, e.g., tanh or ReLU. $W_e$ and $W_p$ are weight matrices, and $\mathbf{b}_e$, $\mathbf{b}_p$ are the biases. $\mathbf{z}_{t_i, u_i}$ is a $|\mathcal{V}|+1$ dimensional logit vector.

The token sequence posterior $P(\mathbf{Y}|\mathbf{X}; \theta^\text{S}_\text{RNNT})$ is computed by summing over the posteriors of all possible blank-augmented token sequences aligned with $\mathbf{Y}$, i.e., $\mathcal{A}(\mathbf{X}, \mathbf{Y})$.
\begin{align}
    & P(\mathbf{Y}|\mathbf{X}; \theta^\text{S}_\text{RNNT}) = \sum_{\mathbf{\tilde{Y}} \in \mathcal{A}(\mathbf{X}, \mathbf{Y})} P(\mathbf{\tilde{Y}}|\mathbf{X}; \theta^\text{S}_\text{RNNT}) \nonumber \\ 
    & \qquad = \sum_{\mathbf{\tilde{Y}} \in \mathcal{A}(\mathbf{X}, \mathbf{Y})} \prod^{T+U}_{i=1} P(\tilde{y}_i|\mathbf{X}_{1:t_i}, \mathbf{Y}_{0:u_{i - 1}}; \theta^\text{S}_\text{RNNT}).
\end{align}

\subsection{Attention-Based Encoder-Decoder (AED)}
For AED, we add a $\texttt{<eos>}$ token to the end of each token sequence to indicate the end of a sentence. The AED model \cite{chorowski2015attention} incorporates an encoder, a decoder and an
attention network. 
The encoder maps a sequence of input speech frames $\mathbf{X}$
into a sequence of hidden representations $\mathbf{H}^{\text{enc}}$.
The attention network 
determines
which encoded features in $\mathbf{H}^\text{enc}$ should be attended to predict
the output label $y_u$ by generating an attention weight for each $\mathbf{h}^\text{enc}_t$ at a decoder step $u$. A context vector $\mathbf{c}_u$ is computed as a linear combination of $\mathbf{H}^{\text{enc}}$ weighted by the attention.
\begin{align}
	\mathbf{a}_u & = \text{AttentionNet}(\mathbf{a}_{u-1}, \mathbf{h}^\text{enc}_t, \mathbf{h}^\text{dec}_u), \\
	\mathbf{c}_{u} & = \sum_{t = 1}^{T} a_{u,t}
		\mathbf{h}^\text{enc}_{t},
	\label{eqn:att}
\end{align}
where $\mathbf{a}_u$ is a vector of attention weights with a dimension of $T$.

At each step $u$, the decoder RNN takes the sum of the previous token embedding 
$\mathbf{e}_{u-1}$ and the context vector $\mathbf{c}_{u-1}$ as the
input to predict the conditional distribution over $\mathcal{V} \cup \texttt{<eos>}$, i.e., 
\begin{align}
       &\mathbf{h}^\text{dec}_u = \text{DecoderRNN}(\mathbf{h}^\text{dec}_{u-1}, \mathbf{e}_{u-1} + \mathbf{c}_{u-1}), \label{eqn:aed_decoder} \\ 
       &\mathbf{z}_u = W_{d}\mathbf{h}^\text{dec}_u +
       \mathbf{b}_d, \label{eqn:aed_logit} \\
       &\left[P(y_u \right.\left.= v| \mathbf{X}, \mathbf{Y}_{0:u-1};\theta^\text{S}_\text{AED})\right]_{v \in
        \mathcal{V} \cup \texttt{<eos>}} \hspace{-3pt} = \text{softmax}(\mathbf{z}_u), \label{eqn:aed_softmax}
\end{align}
where $\mathbf{h}^\text{dec}_u$ is the hidden state of the decoder RNN. $W_d$ and $\mathbf{b}_d$ are weight matrix and bias, respectively.

The token sequence posterior $P(\mathbf{Y}|\mathbf{X}; \theta^\text{S}_\text{AED})$ is computed as
\begin{align}
	P(\mathbf{Y}|\mathbf{X}; \theta^\text{S}_\text{AED}) = \prod_{u = 1}^{U + 1} P(y_u |
		\mathbf{X},\mathbf{Y}_{0:u-1};\theta^\text{S}_\text{AED}).
       \label{eqn:ce}
\end{align}

\section{Related LM Integration Methods}
\label{sec:related}

\subsection{Shallow Fusion}
In Shallow Fusion, a separately-trained external LM is integrated with the E2E model during inference to optimize a log-linear interpolation between the E2E and LM probabilities. The optimal token sequence $\hat{\mathbf{Y}}$ is obtained as follows via a left-to-right beam search
\begin{align}
    \hat{\mathbf{Y}} = \argmax_{\mathbf{Y}} \left[\log P(\mathbf{Y}|\mathbf{X};\theta^\text{S}_\text{E2E}) + \lambda_T \log P(\mathbf{Y};\theta^\text{T}_\text{LM}) \right],
\end{align}
where $P(\mathbf{Y}|\mathbf{X};\theta^\text{S}_\text{E2E})$ is the token sequence posterior of an E2E model trained with source-domain audio-transcript pairs
while $P(\mathbf{Y};\theta^\text{T}_\text{LM})$ is the probability of an external LM trained with target-domain text. $\lambda_T$ is the target-domain LM weight.

\subsection{Density Ratio Method}
In analogy to the traditional DNN-HMM hybrid system, the Density Ratio method \cite{mcdermott2019density} assumes that the token sequence posterior $P(\mathbf{Y}|\mathbf{X};\theta^\text{S}_\text{E2E})$ of a source-domain E2E model is factorizable into a source-domain AM $P(\mathbf{X} | \mathbf{Y}; \theta^\text{S}_\text{AM})$ and a source-domain LM $P(\mathbf{Y};\theta^\text{S}_\text{LM})$ through Bayes' theorem as follows
\begin{align}
    P(\mathbf{Y} | \mathbf{X};\theta^\text{S}_\text{E2E}) = \frac{P(\mathbf{X} | \mathbf{Y}; \theta^\text{S}_\text{AM})P(\mathbf{Y};\theta^\text{S}_\text{LM})}{P(\mathbf{X}; \theta^\text{S}_\text{E2E})}.
    \label{eqn:source_bayes}
\end{align}
Our goal is to estimate the target-domain token sequence posterior $P(\mathbf{Y} | \mathbf{X}; \theta^\text{T}_\text{E2E})$ by using a source-domain E2E model and an external LM trained with target-domain text. Under the same assumption, we have a similar factorization below in the target domain 
\begin{align}
    P(\mathbf{Y} | \mathbf{X}; \theta^\text{T}_\text{E2E}) = \frac{P(\mathbf{X} | \mathbf{Y}; \theta^\text{T}_\text{AM})P(\mathbf{Y};\theta^\text{T}_\text{LM})}{P(\mathbf{X}; \theta^\text{T}_\text{E2E})},
    \label{eqn:target_bayes}
\end{align}
where $P(\mathbf{X} | \mathbf{Y}; \theta^\text{T}_\text{AM})$ is the acoustic likelihood in the target domain, and $P(\mathbf{Y};\theta^\text{T}_\text{LM})$ is the target-domain LM probability.

By assuming domain-invariant acoustic conditions \cite{mcdermott2019density}, i.e., $P(\mathbf{X} | \mathbf{Y}; \theta^\text{S}_\text{AM}) = P(\mathbf{X} | \mathbf{Y}; \theta^\text{T}_\text{AM})$, we derive $P(\mathbf{Y} | \mathbf{X}; \theta^\text{T}_\text{E2E})$ from Eqs. \eqref{eqn:source_bayes} and \eqref{eqn:target_bayes} as follows
\begin{align}
    P(\mathbf{Y} | \mathbf{X}; \theta^\text{T}_\text{E2E}) = P(\mathbf{Y} | \mathbf{X}; \theta^\text{S}_\text{E2E}) \frac {P(\mathbf{Y};\theta^\text{T}_\text{LM})} {P(\mathbf{Y};\theta^\text{S}_\text{LM})} 
    \frac{P(\mathbf{X}; \theta^\text{S}_\text{E2E})}{P(\mathbf{X}; \theta^\text{T}_\text{E2E})},
\end{align}
where $P(\mathbf{X}; \theta^\text{S}_\text{E2E}) / P(\mathbf{X}; \theta^\text{T}_\text{E2E})$ is constant over all hypotheses $\mathbf{Y}$ given $\mathbf{X}$, and $P(\mathbf{Y};\theta^\text{T}_\text{LM})/P(\mathbf{Y};\theta^\text{S}_\text{LM})$ is the \emph{density ratio} between the probabilities of target and source-domain LMs.

During inference, Density Ratio subtracts the log probability of the source-domain LM from the log-linear combination between E2E and the external LM scores, i.e.,
\begin{align}
    \hat{\mathbf{Y}} = \argmax_{\mathbf{Y}} & \left[\log P(\mathbf{Y}|\mathbf{X}; \theta^\text{S}_\text{E2E}) + \lambda_T \log P(\mathbf{Y}; \theta^\text{T}_\text{LM}) \right. \nonumber \\
                     & \left. \quad - \lambda_S \log P(\mathbf{Y}; \theta^\text{S}_\text{LM}) \right],
\end{align}
where $\lambda_S$ is source-domain LM weight.

Note that, in Density Ratio, the source-domain LM $P(\mathbf{Y};\theta^\text{S}_\text{LM})$ is trained with the training transcript of the E2E model, and it has a completely different set of parameters from $\theta^\text{S}_\text{E2E}$.

\section{Internal LM Estimation for E2E Models}
Density Ratio has shown to be more effective than Shallow Fusion for LM integration \cite{mcdermott2019density}. 
However, as in Eqs. \eqref{eqn:source_bayes} and \eqref{eqn:target_bayes}, Density Ratio assumes that the posterior of a source-domain E2E model is decomposable into individual AM and LM with separate parameters like a hybrid system. Strictly speaking, the source-domain E2E posterior should be factorized as follows via Bayes' theorem
\begin{align}
    P(\mathbf{Y} | \mathbf{X};\theta^\text{S}_\text{E2E}) = \frac{P(\mathbf{X} | \mathbf{Y}; \theta^\text{S}_\text{E2E})P(\mathbf{Y};\theta^\text{S}_\text{E2E})}{P(\mathbf{X}; \theta^\text{S}_\text{E2E})},
    \label{eqn:e2e_bayes}
\end{align}
where all factors are conditioned on the same set of E2E parameters $\theta^\text{S}_\text{E2E}$, and $P(\mathbf{Y};\theta^\text{S}_\text{E2E})$ is the \emph{internal LM} of an E2E model.
Given constant acoustic conditions \cite{mcdermott2019density}, i.e., $P(\mathbf{X} | \mathbf{Y}; \theta^\text{S}_\text{E2E}) = P(\mathbf{X} | \mathbf{Y}; \theta^\text{T}_\text{AM})$, the target-domain E2E posterior is computed as follows with Eq. \eqref{eqn:target_bayes}
\begin{align}
    P(\mathbf{Y}|\mathbf{X}; \theta^\text{T}_\text{E2E}) = P(\mathbf{Y} | \mathbf{X}; \theta^\text{S}_\text{E2E}) \frac {P(\mathbf{Y};\theta^\text{T}_\text{LM})} {P(\mathbf{Y};\theta^\text{S}_\text{E2E})} 
    \frac{P(\mathbf{X}; \theta^\text{S}_\text{E2E})}{P(\mathbf{X}; \theta^\text{T}_\text{E2E})}.
\end{align}

During inference, the log probability of the internal LM is subtracted from the log-linear combination between the scores of E2E and external LMs as follow. 
\begin{align}
    \hat{\mathbf{Y}} = \argmax_{\mathbf{Y}} & \left[\log P(\mathbf{Y}|\mathbf{X}; \theta^\text{S}_\text{E2E}) + \lambda_T \log P(\mathbf{Y}; \theta^\text{T}_\text{LM}) \right. \nonumber \\
                     & \left. \quad - \lambda_I \log P(\mathbf{Y}; \theta^\text{S}_\text{E2E}) \right],
\end{align}
where $\lambda_I$ is internal LM weight. We call this LM integration method \emph{internal LM estimation (ILME)}. Compared to Density Ratio, ILME subtracts the log probability of an E2E internal LM parameterized by $\theta^\text{S}_\text{E2E}$ rather than that of a source-domain LM separately-trained with the training transcript of the E2E model.

The key step of ILME is to estimate the internal LM below defined as the token sequence probability distribution an E2E model implicitly learns from the audio-transcript training pairs
\begin{align}
    & P(\mathbf{Y};\theta^\text{S}_\text{E2E}) = \prod^{U}_{u=1}P(y_u|\mathbf{Y}_{0:u-1};\theta^\text{S}_\text{E2E}) \label{eqn:e2e_ilm} \\
    & \qquad = \prod^{U}_{u=1}\prod_{X}P(y_u|\mathbf{X}, \mathbf{Y}_{0:u-1};\theta^\text{S}_\text{E2E}) P(\mathbf{X} | \mathbf{Y}_{0:u-1};\theta^\text{S}_\text{E2E}). \label{eqn:e2e_cond_ilm}
\end{align}
However, the summation over the entire acoustical space in Eq. \eqref{eqn:e2e_cond_ilm} is intractable in practice. To address this, Proposition 1 in \cite[Appdendix A]{variani2020hybrid} was proposed and proved to approximate the internal LM of an HAT model by eliminating the effect of the encoder activations. We name this proposition Joint Softmax Approximation (JSA) in this paper.


JSA suggests that if $\mathbf{f}_t$ is a high-level acoustic representation of the speech feature $\mathbf{x}_t$, $\mathbf{g}_u$ is a language representation of the token $\mathbf{y}_u$, and the output probability of an E2E model can be expressed by $\mathbf{f}_t$ and $\mathbf{g}_u$ in form of $P(y_u| \mathbf{X}, \mathbf{Y}_{0:u-1};\theta^\text{S}_\text{E2E})) = \text{softmax}[J(\mathbf{f}_t + \mathbf{g}_u)]$, where $J$ is any function that satisfies $J(\mathbf{f}_t + \mathbf{g}_u) \approx J(\mathbf{f}_t) + J(\mathbf{g}_u)$, the conditional probability of the E2E internal LM, i.e., $P(y_u|\mathbf{Y}_{0:u-1};\theta^\text{S}_\text{E2E})$, can be approximated as the output of the E2E model at the step $u$ after we zero out the acoustic representation $\mathbf{f}_t$ from the network, i.e., $\text{softmax}[J(\mathbf{g}_u)]$. 

We apply JSA to estimate the internal LMs of the pre-existing most popular E2E models, RNN-T and AED models, in Sections \ref{sec:ilme_rnnt} and \ref{sec:ilme_aed} without any addition training.

\subsection{Internal LM Estimation for RNN-T Model}
\label{sec:ilme_rnnt}
With a softmax token distribution defined in Eq. \eqref{eqn:rnnt_softmax}, the joint network of RNN-T satisfies the conditions of JSA if we re-formulate Eqs. \eqref{eqn:joint} and \eqref{eqn:rnnt_logit} as
\begin{align}
    &\mathbf{f}_{t_i} = W_e\mathbf{h}^{\text{enc}}_{t_i} + \mathbf{b}_e \label{eqn:rnnt_f} \\
    &\mathbf{g}_{u_i} = W_p\mathbf{h}^\text{pred}_{u_i} + \mathbf{b}_p \label{eqn:rnnt_g} \\
    &\mathbf{z}_{t_i, u_i} = J(\mathbf{f}_{t_i} + \mathbf{g}_{u_i}) \label{eqn:rnnt_j} 
\end{align}
where $J(\cdot) = W_j\phi(\cdot) + \mathbf{b}_j$ is a non-linear function followed by an affine transform.
As linear transforms of encoder and prediction network outputs, $\mathbf{f}_{t_i}$, $\mathbf{g}_{u_i}$ are viewed as acoustic and language representations, respectively. By following JSA, we eliminate $\mathbf{f}_{t_i}$ from the joint network and compute the logits as follows
\begin{align}
    \mathbf{z}^\text{ILM}_u & = J(\mathbf{g}_u)=W_j\phi(W_p\mathbf{h}^\text{pred}_u + \mathbf{b}_p) + \mathbf{b}_j \label{eqn:logit_ilm}
\end{align}
Note that, without $\mathbf{f}_{t_i}$, the RNN-T is completely driven by the prediction and joint networks with the token sequence $\mathbf{Y}$ (indexed by $u$) as the only input. No alignment exists between $\mathbf{Y}$ and $\mathbf{X}$ anymore, so we abandon the use of index $i$ for Eqs. \eqref{eqn:logit_ilm} to \eqref{eqn:rnnt_ilm}.

With a designated logit for the blank token $\texttt{<b>}$, $\mathbf{z}^\text{ILM}_u$ has a dimension of $|\mathcal{V}| + 1$. However, we want to estimate the internal LM of the \emph{non-blank} token sequences $\mathbf{Y}$.
Therefore, we remove the blank logit from the vector $\mathbf{z}^\text{ILM}_u$ to construct a new logit vector $\mathbf{z}^\text{ILM, NB}_u$ of dimension $|\mathcal{V}|$, and apply softmax over $\mathbf{z}^\text{ILM, NB}_{u_i}$ to compute the internal LM conditional distribution over $|\mathcal{V}|$ non-blank tokens below
\begin{align}
    & P(y_u|\mathbf{Y}_{0:u-1}; \theta^\text{S}_\text{RNNT}) = \text{softmax}(\mathbf{z}^\text{ILM, NB}_{u}) \label{eqn:rnnt_cond_ilm}
\end{align}
With Eq. \eqref{eqn:e2e_ilm}, the RNN-T internal LM is estimated as 
\begin{align}
    \log P_\text{ILM-RNNT}(\mathbf{Y}) = \sum^{U}_{u=1}\log P(y_u|\mathbf{Y}_{0:u-1};\theta^\text{S}_\text{RNNT})  \label{eqn:rnnt_ilm}
\end{align}

The procedure of the ILME method for LM integration with RNN-T model is the following
\begin{enumerate}
    \item Train a \emph{standard} RNN-T model with source-domain audio-transcript training pairs.
    \item Train an external LM with target-domain text-only data.
    \item During inference, at each step of the beam search, estimate internal LM score of the next \emph{non-blank} candidate token in $\mathcal{V}$ with Eq. \eqref{eqn:rnnt_cond_ilm}, and subtract it from the log-linear interpolation between RNN-T and external LM scores of the same token given a partial hypothesis in the beam using Eq. \eqref{eqn:rnnt_score}. Use only the RNN-T score in Eq. \eqref{eqn:rnnt_softmax} for the candidate token $\texttt{<b>}$.
\end{enumerate}
\begin{align}
    & \text{Score}(\tilde{y}_{u_i} | \mathbf{X}_{1:t_{i-1}}, \mathbf{Y}_{0:u_{i-1}}) \nonumber \\
    & \qquad \quad = \log P(\tilde{y}_i|\mathbf{X}_{1:t_i}, \mathbf{Y}_{0:u_{i - 1}}; \theta^\text{S}_\text{RNNT})  \nonumber \\
    & \qquad \qquad \qquad + \lambda_T \log P(\tilde{y}_i|\mathbf{Y}_{0:u_{i - 1}}; \theta^\text{T}_\text{LM}) \nonumber \\
    & \qquad \qquad \qquad \qquad - \lambda_I \log P(\tilde{y}_{u_i}|\mathbf{Y}_{0:u_{i-1}};\theta^\text{S}_\text{RNNT})
    \label{eqn:rnnt_score}
\end{align}



\subsection{Internal LM Estimation for AED Model}
\label{sec:ilme_aed}
The output distribution of AED is defined by a softmax function in Eq. \eqref{eqn:aed_softmax}. As a special case of JSA when $t$ and $u$ are synchronized (i.e., $t=u$), the decoder of AED also satisfies all its conditions once we re-formulate Eqs. \eqref{eqn:aed_decoder} and \eqref{eqn:aed_logit} as
\begin{align}
    &\mathbf{f}_{u} = \mathbf{c}_{u-1}, \label{eqn:aed_f} \\
    &\mathbf{g}_{u} = \mathbf{e}_{u-1}, \label{eqn:aed_g} \\
    &\mathbf{z}_u = J(\mathbf{f}_u + \mathbf{g}_u) \label{eqn:aed_j} 
\end{align}
where $J(\cdot) = W_d \text{DecoderRNN}(\mathbf{h}^\text{dec}_{u-1}, \cdot) + \mathbf{b}_d$ is a series of linear and non-linear functions. We view $\mathbf{f}_u$, a linear combination of encoder output, as the acoustic representation, and $\mathbf{g}_u$, the token embedding, as the language representation.

JSA allows us to remove $\mathbf{f}_u$, i.e., the context vector $\mathbf{c}_{u}$, from the decoder, and compute the conditional probability of the AED internal LM by applying a softmax over logits output as follows
\begin{align}
    & P(y_u|\mathbf{Y}_{0:u-1}; \theta^\text{S}_\text{AED}) = \text{softmax}\left[J(\mathbf{g}_u)\right]  \nonumber \\
    & = \text{softmax}\left[W_d \cdot \text{DecoderRNN}(\mathbf{h}^\text{dec}_{u-1}, \mathbf{e}_{u-1}) + \mathbf{b}_d \right] \label{eqn:aed_cond_ilm}
\end{align}
Note that, without $\mathbf{f}_{t_i}$, AED is entirely driven by the decoder with the token sequence $\mathbf{Y}$ as the only input, acting exactly the same as an RNN-LM. 
With Eq. \eqref{eqn:e2e_ilm}, the AED internal LM is estimated as 
\begin{align}
    \log P_\text{ILM-AED}(\mathbf{Y}) = \sum^{U+1}_{u=1}\log P(y_u|\mathbf{Y}_{0:u-1};\theta^\text{S}_\text{AED})  \label{eqn:aed_ilm}
\end{align}

The procedure of the ILME method for LM integration with AED model is the following
\begin{enumerate}
    \item Train a \emph{standard} AED model with the source-domain audio-transcript training pairs.
    \item Train an external LM with target-domain text-only data.
    \item During inference, at each step of the beam search, estimate internal LM score of the next candidate token in $\mathcal{V}$ with Eq. \eqref{eqn:aed_cond_ilm}, and subtract it from the log-linear interpolation between AED and external LM scores of the same token given a partial hypothesis in the beam using Eq. \eqref{eqn:aed_score}.
\end{enumerate}
\begin{align}
    & \text{Score}(y_{u} | \mathbf{X}, \mathbf{Y}_{0:u-1}) 
    = \log P(y_u|\mathbf{X}, \mathbf{Y}_{0:u - 1}; \theta^\text{S}_\text{AED})  \nonumber \\
    & \quad \qquad \qquad \qquad \qquad + \lambda_T \log P(y_u|\mathbf{Y}_{0:u - 1}; \theta^\text{T}_\text{LM}) \nonumber \\
    & \quad \qquad \qquad \qquad \qquad - \lambda_I \log P(y_u|\mathbf{Y}_{0:u-1};\theta^\text{S}_\text{AED})
    \label{eqn:aed_score}
\end{align}

\section{Experiment}


In this work, we perform both cross-domain and intra-domain evaluations with different LM integration methods. For all evaluations in this paper, we perform beam search inference with a beam size of 25. All E2E models and RNN-LMs predict word-piece output tokens. We generate 3999 word pieces as the token vocabulary $\mathcal{V}$ by running byte-pair encoding \cite{sennrich2015neural} on acoustic training text, and tokenize all the training transcripts of E2E models and the training text of LMs with these word pieces. We use PyTorch toolkit \cite{paszke2017automatic} for all the experiments.
 
\subsection{E2E Models}
\label{sec:exp_e2e}

\subsubsection{Training data}
We train an RNN-T and an AED with the same 30 thousand (K) hours of anonymized and transcribed data from Microsoft services, including desktop and mobile voice search, short message dictation, command and control, and one-on-one and multi-party conversations recorded in both close-talk and far-field conditions.
We extract 80-dimensional log Mel
filter bank features from the speech signal in both the training and test sets every 10 ms over a 25 ms window. We stack 3 consecutive frames and stride the stacked frame by 30 ms, to form a sequence of 240-dimensional input speech features.  

\subsubsection{RNN-T Model}
We train a 30K-hour RNN-T model predicting word pieces for E2E ASR. The encoder is a uni-directional long short-term memory (LSTM)
\cite{sak2014long, meng2017deep, erdogan2016multi} with 6 hidden layers, each with 1024 hidden units. Each word-piece is represented
by a 1024-dimensional embedding vector. The prediction network is a
uni-directional LSTM with 2 hidden layers, each with 1024 hidden units. The outputs of encoder and prediction network are projected to 1024-dimensional vectors
after layer normalization \cite{ba2016layer}. 
The joint network has 4000-dimensional output units predicting 3999 word pieces and \texttt{<b>}. Dropout \cite{srivastava2014dropout} with a probability of 0.1 is used in both the encoder and the prediction network. An RNN-T loss is implemented with a memory-efficient forward-backward algorithm and is minimized during training. The RNN-T model has 76M parameters. 

\subsubsection{AED Model}
We train a 30K-hour AED model \cite{chorowski2015attention, meng2019character, gaur2019acoustic} predicting word pieces for E2E ASR. The encoder is a bi-directional LSTM with 6 hidden layers and 780 hidden units in each layer. The hidden vectors from both directions are concatenated at each layer and projected to a 780-dimensional vector before layer normalization \cite{ba2016layer}. Each word-piece is represented
by a 780-dimensional embedding vector. The decoder is a
uni-directional LSTM with 2 hidden layers, each with 1280 hidden units. 
The decoder has 4000-dimensional output units predicting 3999 word pieces and \texttt{<eos>}.
During training, scheduled sampling \cite{bengio2015scheduled} is applied to the decoder at rate of 0.0001. Dropout \cite{srivastava2014dropout} with a probability of 0.1 is used in both encoder and decoder.
A label-smoothed cross-entropy \cite{chorowski2016towards} loss is minimized during training.  The AED model has 97M parameters.




\subsection{Cross-Domain Evaluation}
\label{sec:cross_eval}
We evaluate a source-domain E2E model on target-domain test data by integrating an external LM trained with the text-only data in the target domain. We use the RNN-T or AED trained with 30K-hour multi-conditional data in Section \ref{sec:exp_e2e} as the source-domain E2E model, and define LibriSpeech data \cite{panayotov2015librispeech}, read English speech based on LibriVox's audio books, as the target domain. 
The LibriSpeech corpus consists of 960 hours of transcribed training speech and additional 813M words of text-only data collected from 14.5K books. Note that the source-domain 30K-hour training data includes neither the LibriSpeech data nor any read speech from public books. The short message dictation in the 30K-hour data has a very different style from audio books.

We train a target-domain word-piece LSTM-LM with a combination of the 9.4M-word transcript of the 960-hour training speech and the 813M-word text in LibriSpeech. 
The LSTM-LM has 2 hidden layers with 2048 hidden units for each layer. The top hidden layer is first projected to a 512-dimensional embedding, and is then mapped to 4000 output units predicting 3999 word pieces and \texttt{<eos>}. Each token is represented by a 512-dimensional embedding. 
The parameters of input and output embeddings are tied. The LSTM-LM is trained using noise-contrastive estimation loss \cite{mnih2012fast}.
The LibriSpeech LSTM-LM has 58M parameters in total. For Density Ratio, we train a source-domain word-piece LSTM-LM with 2 hidden layers and 2048 hidden units at each layer using the 271M-word transcript of the 30K data. The source-domain LM consists of 57M parameters.

We first evaluate the 30K-hour E2E models on a target-domain test set, LibriSpeech ``test-clean'' with 53K words by using LibriSpeech ``dev-clean'' with 54K words as the validation set for tuning the LM weights. The results are shown in Table \ref{table:wer_libri_clean}. All three LM integration methods show relative WER reductions in the range of 16.1\%-29.1\% from the baseline RNN-T. ILME performs the best achieving 15.5\% and 5.6\% relative WER reductions compared to shallow fusion and density ratio, respectively. The results for AED are similar to that of RNN-T with ILME performing the best, achieving 43.4\%, 8.6\% and 4.3\% relative WER reductions from AED baseline, Shallow Fusion and Density Ratio, respectively.


\begin{table}[h]
\centering
\setlength{\tabcolsep}{4.4pt}
\begin{tabular}[c]{c|c|c|c|c|c|c|c}
	\hline
	\hline
	\multirow{2}{*}{\begin{tabular}{@{}c@{}} E2E \\ Model \end{tabular}} &
	\multirow{2}{*}{\begin{tabular}{@{}c@{}} Method \end{tabular}} & 
	\multirow{2}{*}{\begin{tabular}{@{}c@{}} Params \end{tabular}} & 
	\multirow{2}{*}{\begin{tabular}{@{}c@{}} $\lambda_T$ \end{tabular}} & \multirow{2}{*}{\begin{tabular}{@{}c@{}} $\mu$ \end{tabular}} & 
	\multicolumn{2}{c|}{WER} &
	\multirow{2}{*}{\begin{tabular}{@{}c@{}} Test \\ WERR \end{tabular}} \\
	\hhline{~~~~~--~}
	& & & & & Dev & Test & \\
	\hline
	\multirow{4}{*}{\begin{tabular}{@{}c@{}} RNN-T
		\end{tabular}} & BS & 76M & - & - & 9.27 & 8.97 & - \\
	\hhline{~-------}
	& SF & 134M & 0.12 & 0.00 & 7.44 & 7.53 & 16.1 \\
	\hhline{~-------}
	& DR & 191M & 0.17 & 0.07 & 6.80 & 6.74 & 24.9 \\
	\hhline{~-------}
	& ILME & 134M & 0.30 & 0.14 & 6.41 & 6.36 & 29.1 \\
	\hline
	\hline
	\multirow{4}{*}{\begin{tabular}{@{}c@{}} AED
		\end{tabular}} & BS & 97M & - & - & 8.56 & 8.61 & - \\
	\hhline{~-------}
	& SF & 155M & 0.19	& 0.00 & 5.00 & 5.33 & 38.1 \\
	\hhline{~-------}
	& DR & 212M & 0.15 & 0.03 & 4.74 & 5.09 & 40.9 \\
	\hhline{~-------}
	& ILME & 155M & 0.18 & 0.10 & 4.42 & 4.87 & 43.4 \\
	\hline
	\hline
\end{tabular}
\vspace{-3pt}
  \caption{WERs (\%) of 30k-hour E2E models baseline (BS), Shallow Fusion (SF), Density Ratio (DR) and the proposed ILME methods on the \textbf{out-of-domain LibriSpeech dev-clean} (Dev) and \textbf{test-clean} (Test) sets. The external LM is trained with text-only data in LibriSpeech. LM weight $\mu=\lambda_S$ for DR, and $\mu=\lambda_I$ for ILME. WERR is the relative WER reduction from BS.}
\label{table:wer_libri_clean}
\end{table}



Then we evaluate the 30K-hour E2E models on another target-domain test set, LibriSpeech ``test-other'' with 52K words by using LibriSpeech ``dev-other'' with 51K words as the validation set. The results are shown in Table \ref{table:wer_libri_other}. All three LM integration methods show WER reductions in the range of 6.7\%-15.9\% relative to the baseline RNN-T. ILME performs the best achieving 9.9\% and 5.9\% relative WER reductions compared to shallow fusion and density ratio, respectively. The results for AED are similar to that of RNN-T with ILME performing the best, achieving 44.1\%, 8.6\% and 4.3\% relative WER reductions from AED baseline, Shallow Fusion and Density Ratio, respectively.

\begin{table}[h]
\centering
\setlength{\tabcolsep}{3.8pt}
\begin{tabular}[c]{c|c|c|c|c|c|c|c}
	\hline
	\hline
	\multirow{2}{*}{\begin{tabular}{@{}c@{}} E2E \\ Model \end{tabular}} &
	\multirow{2}{*}{\begin{tabular}{@{}c@{}} Method \end{tabular}} & 
	\multirow{2}{*}{\begin{tabular}{@{}c@{}} Params \end{tabular}} & 
	\multirow{2}{*}{\begin{tabular}{@{}c@{}} $\lambda_T$ \end{tabular}} & \multirow{2}{*}{\begin{tabular}{@{}c@{}} $\mu$ \end{tabular}} & 
	\multicolumn{2}{c|}{WER} &
	\multirow{2}{*}{\begin{tabular}{@{}c@{}} Test \\ WERR \end{tabular}} \\
	\hhline{~~~~~--~}
	& & & & & Dev & Test & \\
	\hline
	\multirow{4}{*}{\begin{tabular}{@{}c@{}} RNN-T \end{tabular}} 
	& BS & 76M & - & - & 20.03 & 20.23 & - \\
	\hhline{~-------}
	& SF & 134M & 0.07 & 0.00 & 18.37 & 18.88 & 6.7 \\
	\hhline{~-------}
	& DR & 191M & 0.20 & 0.12 & 16.14 & 18.07 & 10.7 \\
	\hhline{~-------}
	& ILME & 134M & 0.24 & 0.12 & 15.48 & 17.01 & 15.9 \\
	\hline
	\hline
	\multirow{4}{*}{\begin{tabular}{@{}c@{}} AED \end{tabular}}
	& BS & 97M & - & - & 18.10 & 22.04 & - \\
	\hhline{~-------}
	& SF & 155M & 0.10 & 0.00 & 12.84 & 13.39 & 39.2 \\
	\hhline{~-------}
	& DR & 212M & 0.12 & 0.02 & 12.22 & 12.86 & 41.7 \\
	\hhline{~-------}
	& ILME & 155M & 0.13 & 0.10 & 11.72 & 12.31 & 44.1 \\
	\hline
	\hline
\end{tabular}
\vspace{-3pt}
  \caption{WERs (\%) of 30k-hour E2E models baseline (BS), Shallow Fusion (SF), Density Ratio (DR) and the proposed ILME methods on the \textbf{out-of-domain LibriSpeech dev-other} (Dev) and \textbf{test-other} (Test) set. The external LM is trained with text-only data in LibriSpeech. LM weight $\mu=\lambda_S$ for DR, and $\mu=\lambda_I$ for ILME. WERR is the relative WER reduction from BS.}
\label{table:wer_libri_other}
\vspace{-8pt}
\end{table}


\subsection{Intra-Domain Evaluation}
\label{sec:intra_eval}
We evaluate a multi-conditional E2E model on intra-domain test data by integrating a strong external LM trained with a large amount of multi-domain text-only data. We use the 30K-hour multi-conditional RNN-T or AED model in Section \ref{sec:exp_e2e} for evaluation. 

We train a strong multi-domain word-piece LSTM-LM on a 2B-word text corpus, consisting primarily of conversational data such as talks, interviews, and meeting transcripts. We augmented it with anonymized data from relevant Microsoft services such as short message dictation. With a size of 58M parameters, the 2B-word LSTM-LM has 2 hidden layers with 2048 units for each layer. For Density Ratio, we use the same source-domain LSTM-LM in Section \ref{sec:cross_eval}.


We first evaluate the 30K-hour E2E models on the in-domain in-house dictation test set. With 15K words in total, in-house dictation test set consists of dictation speech collected from keyboard voice input. It has a similar style as the dictation data in 30K corpus and is thus considered as in-domain test data.
We use email dictation data with 9K words as the validation set for tuning the LM weights.
The results are shown in Table \ref{table:wer_dictation}. All three LM integration methods show relative WER reductions in the range of 2.4\%-9.0\% from the baseline RNN-T. ILME performs the best achieving 6.8\% and 6.0\% relative WER reductions compared to shallow fusion and density ratio, respectively. The results for AED are similar to that of RNN-T with ILME performing the best, achieving 12.2\%, 4.6\% and 4.1\% relative WER reductions from AED baseline, Shallow Fusion and Density Ratio, respectively.

\begin{table}[h]
\centering
\setlength{\tabcolsep}{5pt}
\begin{tabular}[c]{c|c|c|c|c|c|c}
	\hline
	\hline
	\multirow{2}{*}{\begin{tabular}{@{}c@{}} E2E \\ Model \end{tabular}} &
	\multirow{2}{*}{\begin{tabular}{@{}c@{}} Method \end{tabular}} & 
	\multirow{2}{*}{\begin{tabular}{@{}c@{}} Params \end{tabular}} & 
	\multirow{2}{*}{\begin{tabular}{@{}c@{}} $\lambda_T$ \end{tabular}} &
	\multirow{2}{*}{\begin{tabular}{@{}c@{}} $\mu$ \end{tabular}} &
	\multicolumn{2}{c}{Dictation} \\
	\hhline{~~~~~--}
	& & & & & WER & WERR \\
	\hline
	\multirow{4}{*}{\begin{tabular}{@{}c@{}} RNN-T
		\end{tabular}} & BS & 76M & - & - & 16.16 & - \\
	\hhline{~------}
	& SF & 134M & 0.03 & 0.00 & 15.77 & 2.4 \\
	\hhline{~------}
	& DR & 191M & 0.21 & 0.19 & 15.64 & 3.2 \\
	\hhline{~------}
	& ILME & 134M & 0.26 & 0.20 & 14.70 & 9.0 \\
	\hline
	\hline
	\multirow{4}{*}{\begin{tabular}{@{}c@{}} AED
		\end{tabular}} & BS & 97M & - & - & 14.08 & - \\
	\hhline{~------}
	& SF & 155M & 0.09 & 0.00 & 12.96 & 8.0 \\
	\hhline{~------}
	& DR & 212M & 0.09 & 0.05 & 12.89 & 8.5 \\
	\hhline{~------}
	& ILME & 155M & 0.13 & 0.08 & 12.36 & 12.2 \\
	\hline
	\hline
\end{tabular}
\vspace{-3pt}
  \caption{WERs (\%) of 30k-hour E2E models baseline (BS), Shallow Fusion (SF), Density Ratio (DR) and the proposed ILME methods on the \textbf{in-domain} \textbf{in-house dictation} test set. The external LM is trained with 2B-word multi-domain text. LM weight $\mu=\lambda_S$ for DR, and $\mu=\lambda_I$ for ILME. WERR is the relative WER reduction.}
\label{table:wer_dictation}
\vspace{-5pt}
\end{table}


Then we evaluate the 30K-hour E2E models on a in-domain in-house meeting test set. With 5K words in total, in-house meeting set consists of conversation speech collected from real meetings. It has a similar style as the conversation data in 30K-hour corpus, and is thus considered as in-domain test data.
Simulated meeting data with 22K words consisting of read meeting transcriptions is used as the validation set.
The results are shown in Table \ref{table:wer_meeting}. All three LM integration methods show WER reductions in the range of 0.2\%-3.2\% relative to the baseline RNN-T. ILME performs the best achieving 2.4\% and 3.1\% relative WER reductions compared to shallow fusion and density ratio, respectively. The results for AED are similar to that of RNN-T with ILME performing the best, achieving 6.2\%, 3.3\% and 3.2\% relative WER reductions from AED baseline, Shallow Fusion and Density Ratio, respectively.

\begin{table}[h]
\centering
\setlength{\tabcolsep}{5pt}
\begin{tabular}[c]{c|c|c|c|c|c|c}
	\hline
	\hline
	\multirow{2}{*}{\begin{tabular}{@{}c@{}} E2E \\ Model \end{tabular}} &
	\multirow{2}{*}{\begin{tabular}{@{}c@{}} Method \end{tabular}} & 
	\multirow{2}{*}{\begin{tabular}{@{}c@{}} Params \end{tabular}} & 
	\multirow{2}{*}{\begin{tabular}{@{}c@{}} $\lambda_T$ \end{tabular}} &
	\multirow{2}{*}{\begin{tabular}{@{}c@{}} $\mu$ \end{tabular}} & 
	\multicolumn{2}{c}{Meeting} \\
	\hhline{~~~~~--}
	& & & & & WER & WERR \\
	\hline
	\multirow{4}{*}{\begin{tabular}{@{}c@{}} RNN-T
		\end{tabular}} & BS & 76M & - & - & 20.64 & - \\
	\hhline{~------}
	& SF & 134M & 0.02 & 0.00 & 20.47 & 0.8 \\
	\hhline{~------}
	& DR & 191M & 0.03 & 0.02 & 20.60 & 0.2 \\
	\hhline{~------}
	& ILME & 134M & 0.08 & 0.03 & 19.97 & 3.2 \\
	\hline
	\hline
	\multirow{4}{*}{\begin{tabular}{@{}c@{}} AED
		\end{tabular}} 
	& BS & 97M & - & - & 19.46 & - \\
	\hhline{~------}
	& SF & 155M & 0.09 & 0.00 & 18.87 & 3.0 \\
	\hhline{~------}
	& DR & 212M & 0.11 & 0.03 & 18.85 & 3.1 \\
	\hhline{~------}
	& ILME & 155M & 0.12 & 0.08 & 18.25 & 6.2 \\
	\hline
	\hline
\end{tabular}
\vspace{-3pt}
  \caption{WERs (\%) of 30k-hour E2E models baseline (BS), Shallow Fusion (SF), Density Ratio (DR) and the proposed ILME methods on the \textbf{in-domain} \textbf{in-house meeting} test set. The external LM is trained with 2B-word multi-domain text. LM weight $\mu=\lambda_S$ for DR, and $\mu=\lambda_I$ for ILME. WERR is the relative WER reduction from BS.}
\label{table:wer_meeting}
\vspace{-8pt}
\end{table}

\subsection{Result Analysis}
By summarizing the results in Sections \ref{sec:cross_eval} and \ref{sec:intra_eval}, we have the following observations for all test sets in both cross-domain and intra-domain evaluations.
For both E2E models, ILME consistently reduces the WER of Shallow Fusion by 8.1\%-15.5\% relatively in cross-domain evaluation, and 2.4\%-6.8\% relatively in intra-domain evaluation. ILME also consistently outperforms Density Ratio in terms of lower WER, and achieves the goal with 29.8\% and 26.8\% fewer run-time parameters with RNN-T and AED, respectively, because ILME does not require an additional source-domain LSTM-LM during beam search. 

We also notice that the baseline WER for AED is lower than RNN-T because AED has a bi-directional encoder and has more parameters than RNN-T.
With larger relative WER reductions from the AED baseline, all LM integration methods are more effective for AED than for RNN-T. However, the benefit of subtracting internal LM is larger for RNN-T than for AED because ILME gets greater relative WER reductions with RNN-T than with AED from Shallow Fusion. We also notice that, for all three LM integration methods, most of the gains they obtain are transferable from the dev sets to the test sets with both E2E models, showing that they all have good generalization capability. 

By comparing cross-domain and intra-domain evaluations, we notice that all LM integration methods obtain much higher relative WER reductions from the baseline E2E model in cross-domain evaluation (LibriSpeech test-clean, test-other sets) than in intra-domain evaluation (in-house dictation, in-house meeting). This is because, in intra-domain evaluation, the E2E training data has been exposed to data with similar language styles as in testing, diminishing the effectiveness of an external LM. However, even on in-domain test data, ILME can still get 3.2\%-9.0\% relative WER reductions from the baseline RNN-T, significantly larger than the 0.2\%-3.2\% relative WER reductions obtained by Shallow Fusion and Density Ratio. This shows that the subtraction of internal LM indeed enables a more effective integration of the external LM, and thus a better use of the abundant external knowledge from text-only data.

\subsection{Internal LM Perplexity}

\begin{figure}[htb]
\vspace{-2pt}
\begin{minipage}[b]{.48\linewidth}
  \centering
  \centerline{\includegraphics[width=4.48cm]{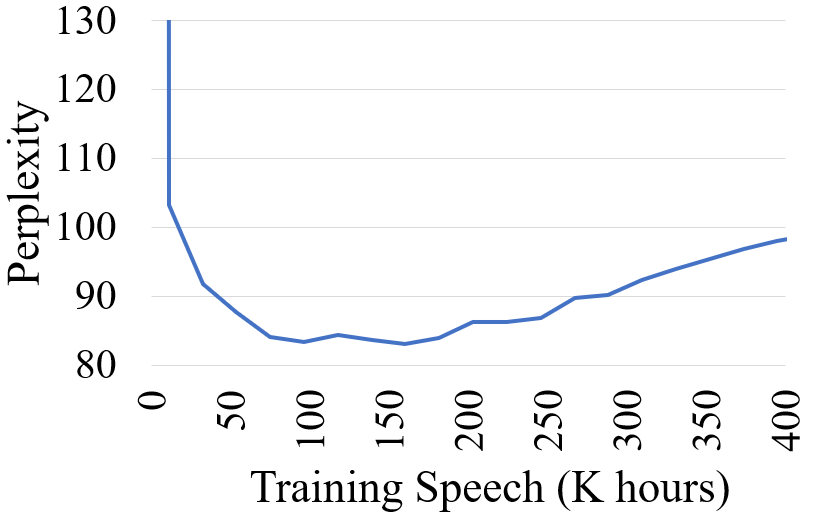}}
  \centerline{(a) RNN-T Internal LM}\medskip
\end{minipage}
\hfill
\begin{minipage}[b]{0.48\linewidth}
  \centering
  \centerline{\includegraphics[width=4.48cm]{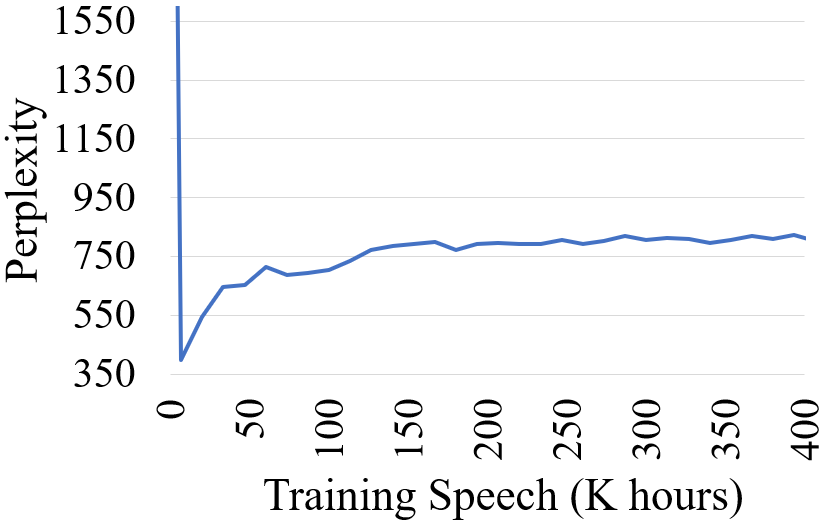}}
  \centerline{(b) AED Internal LM}\medskip
\end{minipage}
\vspace{-13pt}
\caption{The internal LM perplexity of RNN-T and AED as a function of thousand (K) hours of training speech. The perplexity is evaluated on the validation set of the 30K-hour E2E training data.}
\label{fig:ilm_ppl}
\vspace{-2pt}
\end{figure}

We evaluate the internal LM perplexities (ILM-PPLs) of AED and RNN-T models at different epochs on the validation set of 30K-hour training data which is drawn from the same distribution as the training data. In Fig. \ref{fig:ilm_ppl}, both ILM-PPLs first decrease and then increase as the E2E models get trained with more data. Both ILM-PPLs fall sharply at initial epochs, but the one of AED begins to increase and then gradually converge at much earlier training stages than that of RNN-T. 
The significantly larger ILM-PPL of AED than RNN-T suggests that 
in AED, the decoder needs to work together with the encoder to perform the language modeling. The decoder does not function well as a standalone LM without the encoder input. However, in RNN-T, the prediction network is trained to act like an individual LM and is later combined with the encoder by the joint network.

The ILM-PPLs of the RNN-T and AED used in Sections \ref{sec:cross_eval} and \ref{sec:intra_eval} are 99.4 and 796.7, respectively, while the source-domain LM trained with 30K-hour training transcript has a PPL of 30.1 on the same data. Given the maximum perplexity of 4000 under uniform distribution over tokens, the internal LM indeed learns, to some extent, the distribution of token sequences through the E2E training.


\section{Conclusion}
We propose an internal LM estimation method to effectively integrate an external LM with all pre-existing E2E models without any additional training.
The internal LM is approximated by the output of an E2E model when zeroing out its acoustic components. During inference, the internal LM score is subtracted from the log-linear combination of E2E model and LM scores.
We evaluate 30K-hour RNN-T and AED models on cross-domain LibriSpeech test sets and intra-domain Microsoft production test sets with different LM integration methods. ILME reduces the WER of Shallow Fusion by 8.1\%-15.5\% relatively in cross-domain evaluations, and 2.4\%-6.8\% relatively in intra-domain evaluations. Further, ILME consistently outperforms Density Ratio in terms of better WER despite having 26.8\%-29.8\% fewer run-time parameters. 

\vfill
\pagebreak

\bibliographystyle{IEEEbib}
\bibliography{strings,refs}

\end{document}